\documentclass[11pt]{article}

\usepackage[left=2cm, right=2cm, top=2cm, bottom=2cm]{geometry}
\usepackage{setspace}
\usepackage{indentfirst}
\usepackage{tabularx}

\usepackage[T1]{fontenc}
\usepackage{graphicx}
\usepackage{titlesec}
\usepackage{titling}

\usepackage{amsmath}
\usepackage{amssymb}
\usepackage{booktabs}
\usepackage{makecell}
\usepackage{multirow}
\usepackage{longtable}
\usepackage{array}
\usepackage{enumitem}

\usepackage[hyphens]{url}
\usepackage{hyperref}
\hypersetup{
    colorlinks=true,
    linkcolor=blue,
    urlcolor=blue,
    breaklinks=true
}

\pretitle{%
  \begin{center}%
  \rule{\linewidth}{0.6pt}\\[1.5em]
  \LARGE\bfseries
}
\posttitle{%
  \\[0.75em]
  \rule{\linewidth}{0.6pt}
  \end{center}
  \vspace{1.5em}
}

\preauthor{\begin{center}\large}
\postauthor{\end{center}\vspace{0pt}}

\predate{}
\postdate{}
\date{}  

\begin{document}

\title{Non-Linear Determinants of Pedestrian Injury Severity:\\[0.5em]
\Large Evidence from Administrative Data in Great Britain}

\author{
\large Yifei Tong\\[0.25em]
\normalsize Georgetown University\\[0.25em]
\normalsize\texttt{yt583@georgetown.edu}
}

\maketitle

\vspace{-15pt} %
\begin{abstract}
Road traffic injuries remain a leading cause of death and disability, with pedestrians representing one of the most vulnerable road user groups in Great Britain. This study utilizes administrative data from the 2023 STATS19 dataset to investigate the non-linear determinants of pedestrian-involved collisions and injury severity.

The empirical strategy addresses data-quality challenges inherent in administrative safety records, including invalid codes, missing information, and substantial class imbalance. We employ a rigorous preprocessing pipeline, utilizing mode imputation for categorical variables and Synthetic Minority Over-sampling (SMOTE) to address the rarity of severe outcomes. To capture complex interactions and heterogeneity often missed by linear models, we utilize non-parametric ensemble methods (Random Forest and XGBoost). Model interpretability is achieved through Shapley Additive Explanations (SHAP), allowing for the isolation of marginal feature effects.

Our analysis reveals that the number of vehicles, speed limits, lighting, and road surface conditions are the primary predictors of severity, while police attendance and junction characteristics further distinguish severe from non-severe collisions. Spatially, while pedestrian risk is concentrated in dense urban Local Authority Districts (LADs), we identify that certain rural LADs experience disproportionately severe outcomes conditional on a collision occurring. These findings underscore the value of combining spatial analysis with interpretable machine learning to guide geographically targeted speed management and infrastructure investment, and enforcement strategies in transportation and pedestrian safety policy.
\end{abstract}


\section*{1. Introduction}

Road traffic injuries remain a significant public health concern and a leading cause of death and disability worldwide. According to the World Health Organization, approximately 1.3 million people are killed in road traffic crashes each year, with the majority of fatalities occurring in low- and middle-income countries. Within Great Britain, pedestrians represent one of the most vulnerable groups on the road, accounting for 24 percent of all road deaths in 2021 according to the Department for Transport. Understanding the conditions that contribute to heightened pedestrian risk is therefore an important area of inquiry for researchers and policymakers.

Although the United Kingdom has achieved notable progress in reducing road fatalities over the past three decades, the overall volume of traffic casualties remains considerable. Recent policy changes, including the 2022 amendments to the Highway Code that prioritize pedestrians and cyclists at junctions, were enacted with the goal of improving safety for vulnerable road users. However, public awareness, compliance, and enforcement of these changes remain uneven. The Government’s Public Accounts Committee report of November 2023 highlighted limitations in the communication and implementation of the new rules, suggesting that policy revisions alone may be insufficient without broader behavioral and infrastructural support. Existing literature, such as Choudhary et al. (2023), underscores persistent challenges driven by urbanization, increasing vehicle numbers, speeding, and inconsistent adherence to road safety practices.

Against this backdrop, this study examines pedestrian collision risks using the Department for Transport’s 2023 Road Safety Data, combining spatial analysis with predictive modeling to generate a comprehensive understanding of pedestrian safety patterns in the United Kingdom. The spatial component links collision coordinates to Local Authority Districts, allowing identification of geographic concentrations of pedestrian-related incidents and exploration of contextual differences across regions. Complementing this, the predictive modeling framework evaluates a range of environmental, temporal, and road-specific characteristics associated with pedestrian involvement and injury severity. By integrating geographic and model-based perspectives, the study seeks to provide a nuanced assessment of pedestrian safety and to generate evidence that can support policy evaluation, urban planning decisions, and the design of targeted interventions to reduce pedestrian harm.

\section*{2. Methodology}
\subsection*{2.1 Data Source
}

To create a model predicting pedestrian road safety, we utilize the collision dataset from the 2023 Road Accident and Safety Statistics, published by the Department for Transport for the UK Government. The STATS19 dataset, named after the data collection form used by the police, is pivotal for analyzing road traffic collisions and pedestrian safety in Great Britain. It primarily includes statistics from collisions in which at least one person was injured and reported to the police, who subsequently relay the data to the Department for Transport. The dataset comprises 49,316 entries, each representing a unique collision.

While historical statistics have used the terminology "road traffic accidents" in line with the Road Traffic Act, the terms "accidents" and "collisions" are now used interchangeably in the current context of road safety since 2022. For a collision to be included, it must have involved at least one vehicle and resulted in a personal injury. Collisions that do not cause bodily harm, such as “damage-only” incidents, are outside the scope of this dataset and are not estimated.

The dataset captures a wide range of incidents, from accidents involving motor vehicles to those involving pedestrians or pedal cyclists falling from their bicycles on the road, even when no other vehicle or pedestrian was involved. However, the dataset has limitations, as there is no legal obligation for drivers to report a road traffic collision if all parties exchange details, even when injuries occur, which can lead to underreporting. In addition, within the analytical scope of pedestrian safety, focusing solely on vehicle-only accidents may result in underestimation.

Despite this limitation, the road safety dataset remains a valuable resource for predicting pedestrian safety due to its comprehensive collection. The inclusion of pedestrian casualties and the clear definitions of traffic terminologies contribute to the dataset's relevance for analyzing pedestrian safety. Moreover, the technological  advancements in data collection methods, such as injury-based reporting systems and online platforms, greatly enhanced quality and accuracy, according to the Road Casualty Background Quality Report, 2023.

In addition to the collision records, the study incorporates spatial boundary indicators from the Local Authority Districts (December 2023) Boundaries dataset, published by the UK Office for National Statistics. This dataset is provided in GeoPackage format, including polygon geometries and administrative identifiers for all Local Authority Districts across the United Kingdom. Integrating this spatial layer enables each collision to be assigned to a specific administrative area through a spatial join, supporting analysis of geographic patterns in pedestrian-related collisions and facilitating district-level aggregation.

For the forthcoming analysis, Table 1 presents the variables from the original dataset, along with descriptions and notes from the official Background Quality Report to aid understanding. Table 2 introduces the additional column "Observation," which presents the occurrence counts of each target feature before and after revision to assess the need for further imputation or adjustment. The preprocessing steps applied to both the collision data and the spatial boundary data will be discussed in detail in the next section. Note that the minimum, mean, maximum values, and count of observations mentioned in the subsequent analysis are based on the preprocessed data, which may show minor discrepancies compared to the descriptive statistics summary of the raw dataset.

\setlist[itemize]{label={}}

\renewcommand{\arraystretch}{1.1}
\begin{longtable}{>{\raggedright\arraybackslash}m{4.1cm} 
                  >{\raggedright\arraybackslash}m{4.1cm} 
                  >{\raggedright\arraybackslash}m{1cm} 
                  >{\raggedright\arraybackslash}m{0.9cm} 
                  >{\raggedright\arraybackslash}m{0.9cm} 
                  >{\raggedright\arraybackslash}m{5.2cm}}
\caption{Relevant Features Introduction} \label{tab:features_introduction} \\
\toprule
\textbf{Variable Name} & \textbf{Description} & \textbf{Mean} & \textbf{Min} & \textbf{Max} & \textbf{Notes }\\
\midrule
\endfirsthead

\multicolumn{6}{c}%
{{ Table \thetable\ Continued from previous page}} \\
\toprule
\textbf{Variable Name} & \textbf{Description} & \textbf{Mean} & \textbf{Min} & \textbf{Max} & \textbf{Notes} \\
\midrule
\endhead

\midrule
\multicolumn{6}{r}{{Continued on next page}} \\ 
\bottomrule
\endfoot

\bottomrule
\endlastfoot

Number of vehicles & Number of vehicles involved in the accidents & 1.81 & 1 & 17 &  \\
\midrule[0.1pt]

Number of casualties & Number of casualties in the accidents & 1.27 & 1 & 19 &  \\\addlinespace[2pt]
\midrule[0.1pt]

Day of week & The date of the accident & 4.11 & 1 & 7 & Represents the respective day of week \\\addlinespace[5pt]
\midrule[0.1pt]

Nearest hour & Time of the accident (rounded to the nearest hour) & 13.95 & 0 & 24 & The hours are to be entered in the first two boxes and transformed to the nearest hour \\\addlinespace[5pt]
\midrule[0.1pt]

Road type & 
The type of road where the accident took place & 
5.31 & 
1 & 
9 & 
\begin{itemize}[leftmargin=*, nosep] 
    \item 1. Roundabout
    \item 2. One way street
    \item 3. Dual carriageway
    \item 6. Single carriageway
    \item 7. Slip Road
    \item 9. Unknown
\end{itemize} 
\vspace*{0pt} 
\\ 
\addlinespace[5pt] 
\midrule[0.1pt]

Speed limit & The speed limit at the location of the accident 
             & 35.67 & 0 & 70 
             & \\\addlinespace[2pt]
\midrule[0.1pt]

Junction control & 
The type of control mechanism at the junction & 
3.75 & 
1 & 
4 & 
\begin{itemize}[leftmargin=*, nosep]
    \item 1. Authorized person
    \item 2. Auto traffic signal
    \item 3. Stop sign
    \item 4. Uncontrolled
\end{itemize} 
\vspace*{0pt}
\\ \addlinespace[4pt]
\midrule[0.1pt]

Light conditions & 
The illumination level at the accident location & 
1.93 & 
1 & 
7 &
\begin{itemize}[leftmargin=*, nosep]
    \item 1. Daylight
    \item 4. Street lights lit
    \item 5. Street lights unlit
    \item 6. No street lighting
    \item 7. Lighting unknown
\end{itemize}
\vspace*{0pt}
\\ \addlinespace[4pt]
\midrule[0.1pt]

\makecell[tl]{Human control at \\ pedestrian crossing} &
Type of human control at the pedestrian crossing &
0.03 & 
0 & 
2 &
\begin{itemize}[leftmargin=*, nosep]
    \item 0. None within 50 meters
    \item 1. School crossing patrol
    \item 2. Other authorized person
\end{itemize} 
\vspace*{0pt}
\\ \addlinespace[3pt]

Junction detail & 
The type of junction within a 20-meter radius & 
2.40 & 
0 & 
9 & 
\begin{itemize}[leftmargin=*, nosep]
    \item 0. Not near junction
    \item 1. Roundabout
    \item 2. Mini roundabout
    \item 3. T junction
    \item 5. Slip road
    \item 6. Crossroads
    \item 7. More than four arms
    \item 8. Private drive
    \item 9. Other
\end{itemize} 
\vspace*{0pt}
\\ \addlinespace[4pt]
\midrule[0.1pt]

Weather conditions & 
The weather condition at the accident location & 
1.64 & 
1 & 
9 & 
\begin{itemize}[leftmargin=*, nosep]
    \item 1. Fine no high winds
    \item 2. Raining no high winds
    \item 3. Snowing no high winds
    \item 4. Fine high winds
    \item 5. Raining high winds
    \item 6. Snowing high winds
    \item 7. Fog or mist
    \item 8. Other
    \item 9. Unknown
\end{itemize} 
\vspace*{0pt}
\\ \addlinespace[4pt]
\midrule[0.1pt]

Road surface conditions & 
The road surface condition at the accident location &
1.28 & 
1 & 
5 &
\begin{itemize}[leftmargin=*, nosep]
    \item 1. Dry
    \item 2. Wet/Damp
    \item 3. Snow
    \item 4. Frost/Ice
    \item 5. Flooded (over 3 cm deep)
\end{itemize}
\\ \addlinespace[5pt] 
\midrule[0.1pt]

\makecell[tl]{\vspace*{-2pt}Police officer attendance \\ at scene of collision} &
Indicates whether a police officer was present at the scene &
1.53 & 
1 & 
3 &
\begin{itemize}[leftmargin=*, nosep]
    \item 1. Yes
    \item 2. No
    \item 3. Self-reporting form used
\end{itemize}
\vspace*{0pt}
\\ \addlinespace[5pt]

\end{longtable}

\subsubsection*{Target Features and Imputation}

Firstly, we integrate the \texttt{'casualty\_class'} variable from the provisional road casualty statistics report, part of the same 2023 Road Accident and Safety Statistics. This data provides detailed accounts of each accident, including cases where collision references were reported multiple times by different individuals involved. To address this, we create a dummy variable indicating whether any person involved in a collision was a pedestrian. We then merge this variable into the main collision dataset by the reference number, ensuring that each collision is recorded only once. Because the collision file contains one record per incident, there are no duplicate entries for a single collision.

\begin{figure}[ht]
    \hspace*{-2cm} 
    \centering
    \includegraphics[width=12cm]{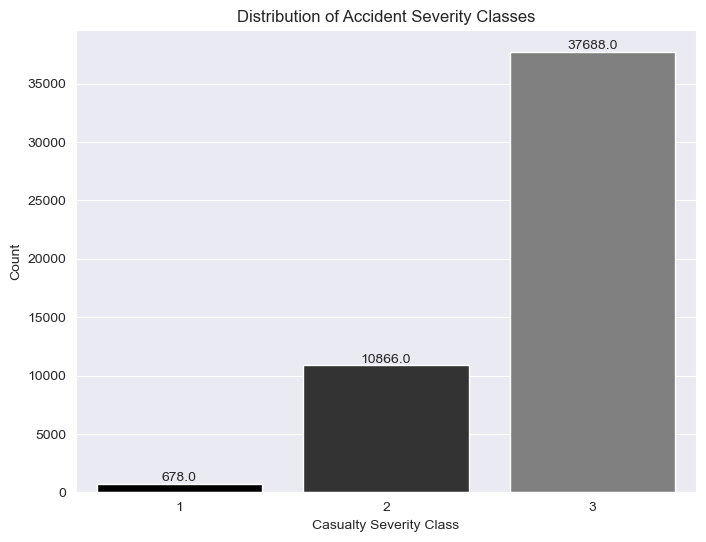}
    \caption{Distribution of Accident Severity Classes}
    \label{fig:sverity_level} 
\end{figure}

From the original distribution plot in Figure~\ref{fig:sverity_level}, we observe a significant class imbalance in our target features, with a disproportionate focus on the majority class. For the Severity Level feature, our particular concern is the minority class, which includes only 678 fatal cases and even fewer (194) pedestrian-related fatalities. This limited data poses challenges for the credibility and robustness of any predictive modeling focused solely on fatal outcomes. To address this, we consider expanding our target feature to include both serious and fatal cases. Combining these categories would not only provide a larger sample size but also maintain a focus on the more serious outcomes, which are of greater interest in safety studies.

\setlist[itemize]{label={}}

\begin{longtable}{>{\raggedright\arraybackslash}m{4cm} 
                >{\raggedright\arraybackslash}m{6.1cm} 
                >{\raggedright\arraybackslash}m{3.2cm} 
                >{\raggedright\arraybackslash}m{4.2cm}}
\caption{Target Feature Introduction} \label{tab:target_feature_introduction} \\
\toprule
\textbf{Variable Name} & \textbf{Description} & \textbf{Observations} & \textbf{Changes for the Project} \\
\midrule
\endfirsthead

\multicolumn{4}{c}%
{{ Table \thetable\ Continued from previous page}} \\
\toprule
\textbf{Variable Name} & \textbf{Description} & \textbf{Observations} & \textbf{Changes for the Project} \\
\midrule
\endhead

\midrule
\multicolumn{4}{r}{{Continued on next page}} \\ 
\bottomrule
\endfoot

\bottomrule
\endlastfoot

casualty\_pedestrian &
Indicates the class of the casualty
(0 = Not pedestrian involved, including driver, rider, or vehicle passenger;
1 = Pedestrian involved) &
\begin{itemize}[leftmargin=*, nosep, before=\vspace{-1.0\baselineskip}, after=\vspace{-1.0\baselineskip}]
  \item Non-pedestrian-involved 40,351
  \item Pedestrian-involved 8,881
\end{itemize} &
Created a dummy variable indicating whether pedestrians were involved \\
\addlinespace[3pt]
\midrule[0.1pt]

casualty\_severity & Indicates the severity level of the casualty (1 = Fatal; 2 = Serious; 3 = Slight) & 
\begin{itemize}[leftmargin=*, nosep, before=\vspace{-1.0\baselineskip}, after=\vspace{-1.0\baselineskip}]
\item Fatal 678
\item Serious 10,866
\item Slight 37,688
\end{itemize} & Recoded from \texttt{casualty\_severity} to flag cases above "serious” \\
\midrule[0.1pt]

casualty\_over\_serious & Indicates whether the severity level of the casualty is more than serious or not & 
\begin{itemize}[leftmargin=*, nosep, before=\vspace{-0.75\baselineskip}, after=\vspace{-0.75\baselineskip}]
\item Not serious 37,668
\item Serious 11,544
\end{itemize} & \\
\midrule[0.1pt]

pedestrian\_over\_serious & Indicates whether the casualty includes pedestrian and is serious & 
\begin{itemize}[leftmargin=*, nosep, before=\vspace{-1.0\baselineskip}, after=\vspace{-1.0\baselineskip}]
\item False 46,409
\item True 2,823
\end{itemize} & Multiplication of \texttt{casualty\_pedestrian} and \texttt{casualty\_over\_serious} to make an interaction term \\
\end{longtable}

\newpage
\subsection*{2.2 Descriptives}

\begin{figure}[ht]
    \hspace*{-2cm} 
    \includegraphics[width=21cm]{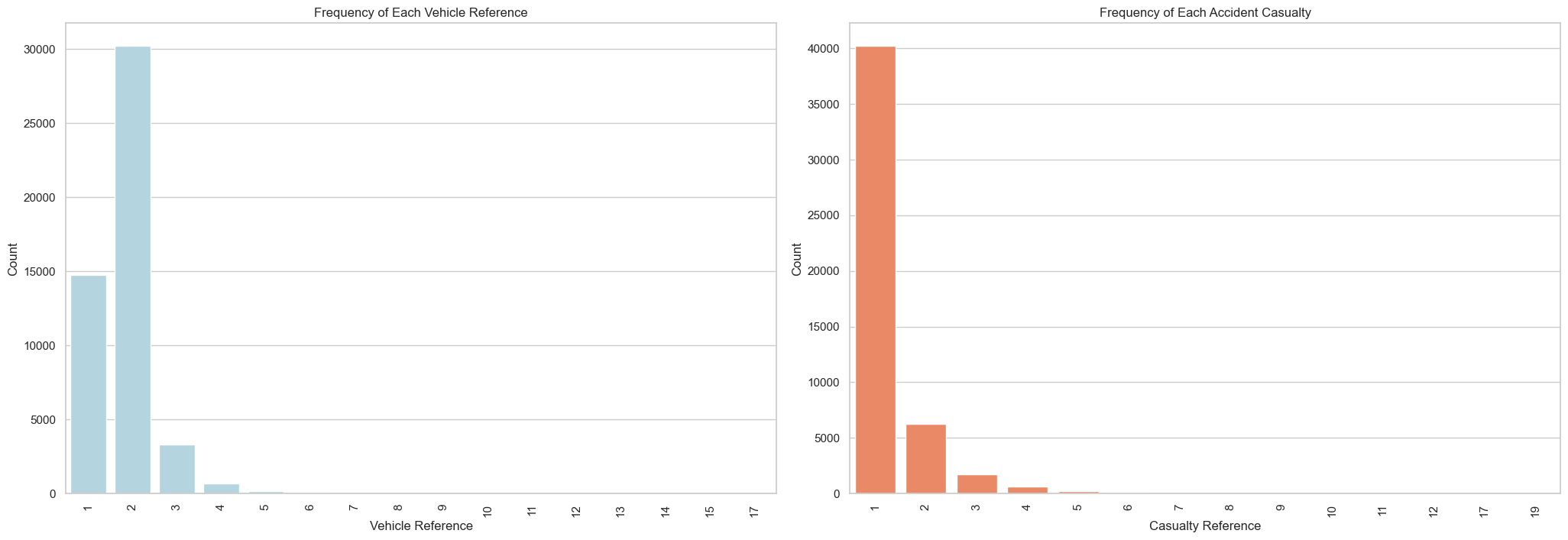}
    \caption{Distribution of Number of Vehicles/Casualties}
    \label{fig:combined}
\end{figure}

Based on the plots and summary statistics before preprocessing, several key insights emerge from the collision data. The mean value of \texttt{number\_of\_vehicles} indicates that, on average, one to two vehicles are involved in each collision. The distribution is highly left-skewed, with most collisions involving only one or two vehicles, reflecting the predominance of smaller-scale incidents over more complex multi-vehicle crashes. Although a few observations show values as high as 17, these are retained in the dataset, as they, while uncommon, fall within acceptable bounds under our criteria for frequency and impact on model performance.

A similar pattern is observed for \texttt{number\_of\_casualties}, which also shows a left-skewed distribution, indicating that most collisions result in a single injury. The distribution peaks sharply at one and declines steeply for higher values with a significant drop-off. The maximum value recorded is 70, representing an extreme case that deviates substantially from the norm. Given the rarity of such incidents and their potential to disproportionately affect model performance, we remove this single outlier from the dataset to ensure a more accurate and representative analysis of typical road traffic collisions.

\begin{figure}[ht]
    \hspace*{-2cm} 
    \includegraphics[width=21cm]{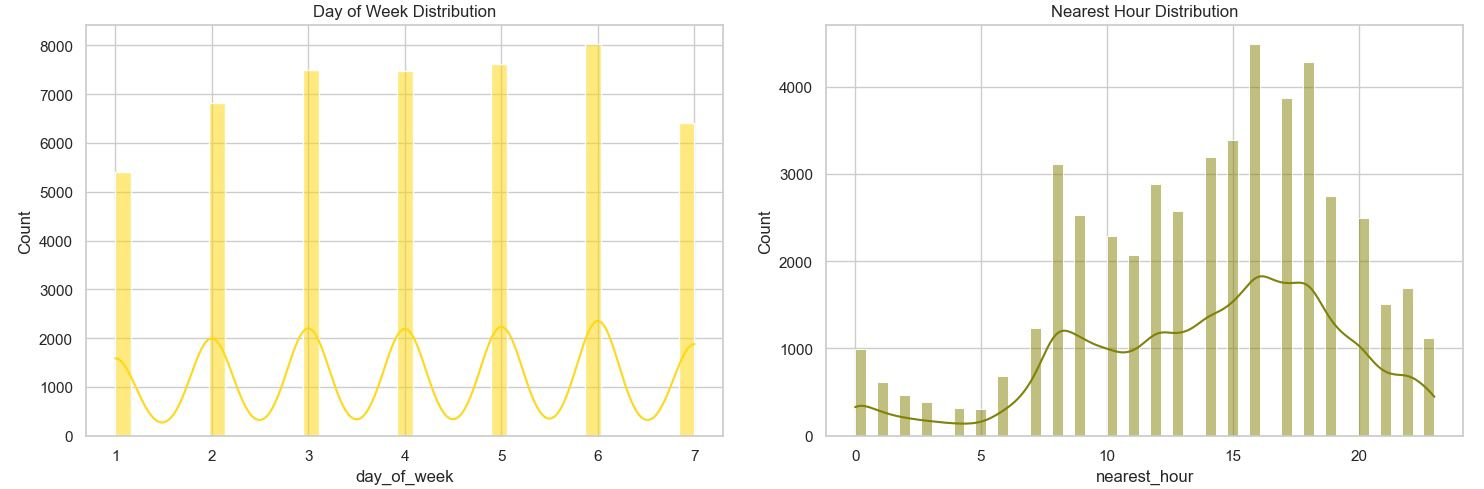}
    \caption{Distribution of Days of Week and Nearest Hour}
    \label{fig:day_hour} 
\end{figure}

The most common speed limit of 30 mph typically corresponds to urban areas with higher traffic density, greater pedestrian activity, and more complex road layouts, all of which can contribute to a higher frequency of accidents. Interestingly, as shown in Figure 3, the plot reveals that accident counts do not peak during extreme weather conditions or periods of poor visibility. Instead, most accidents occur in fine weather, under normal daylight conditions, and on dry roads.
Regarding timing, accident frequency is relatively even across weekdays, with slight reductions on Mondays and Sundays, likely reflecting lower traffic volumes at the beginning and end of the week. A noticeable peak occurs between 10 a.m. and 8 p.m., corresponding to busy daytime hours when traffic volume increases due to commuting, work-related travel, and school runs.

Together with the earlier descriptive analysis, these distributions offer a more comprehensive understanding of the dataset’s underlying composition. Extending previous work, the examination of categorical road characteristics uncovers pronounced skewness in several variables, indicating the predominance of particular structural features in recorded collisions. A salient finding is the high concentration of accidents occurring in locations without traffic signals or authorized personnel. This pattern suggests that environments with limited external regulation may be more susceptible to driver error or misjudgment. In particular, the prevalence of incidents classified as Not at or within 20 meters of a junction in the \texttt{Junction detail} variable indicates that many crashes take place away from intersections. This may imply that straight or uninterrupted road segments carry elevated risk—potentially due to higher travelling speeds, reduced visual cues, or diminished driver vigilance in less complex driving environments.

In addition, \textit{T or staggered junctions} and \textit{Crossroads} emerge as common accident sites. These junction types, characterized by complex traffic flows and multiple conflict points, require careful navigation, which may not always occur, leading to a higher incidence of collisions. The predominance of accidents on single carriageways, where opposing traffic streams are not physically separated, further underscores the need for targeted safety interventions in these areas.\footnote{The remaining distribution plots for defined variables are provided in the Appendix.}

Considering these insights, interventions such as clearer traffic signals, improved road markings, or the installation of traffic lights at key locations could substantially mitigate collision risks. Further quantitative analysis could refine these recommendations by identifying the specific factors that most strongly contribute to elevated accident rates in these areas.

\subsection*{2.3 Preprocessing}

Although no missing values are identified using \texttt{isnull()}, an initial data review reveals that certain variables contain invalid entries represented by the minimum value \texttt{-1} and the maximum value \texttt{99}, both of which are undefined in the dataset’s variable definitions. After reviewing the accompanying documentation, we find no explanation for these extreme codes, prompting further evaluation of how to address them. We first consider excluding all rows containing these invalid values to maintain data integrity; however, 6,312 of the total 49,316 observations (approximately 12.8\%) are affected. Full removal would simplify the dataset but also reduce its size considerably, risking the loss of valuable information and introducing potential bias. Therefore, we adopt an imputation strategy that replaces the invalid values with the mode (most frequent value) of their respective columns. This approach preserves the distributional properties of categorical variables (such as weather and light conditions) without introducing the mathematical bias associated with calculating the median of nominal identifiers.

To better align with our analysis goal of analyzing pedestrian road safety, we modify two original features: \texttt{casualty\_} \texttt{class} and \texttt{casualty\_severity}. We create \texttt{casualty\_pedestrian} to isolate pedestrian cases (1 = pedestrians, 0 = otherwise) and \texttt{casualty\_over\_serious} to categorize accidents as either serious or fatal (1) versus slight (0). The severity of a casualty refers to whether the individual was killed, seriously injured, or slightly injured, and typically reflects the condition of the most severely affected casualty in each incident.

Furthermore, to emphasize severe accidents involving pedestrians, we introduce an interaction term \texttt{pedestrian\_over} \texttt{\_serious} obtained by multiplying two created dummy variables. This variable enables us to specifically target and analyze the most critical subset of accidents, thereby enhancing our ability to identify the most significant factors associated with pedestrian safety in severe incidents.

\subsection*{2.4 Research Design}

In this study, we analyze road safety data from reported collisions in 2023, focusing on pedestrian-related accidents. Before implementing predictive models, we incorporate spatial information by assigning each collision to its corresponding Local Authority District through a point-in-polygon join, enabling district-level aggregation and providing geographic context for the subsequent machine learning analysis. We employ Random Forest and XGBoost models, selected for their complementary strengths: Random Forest for its robustness in capturing complex nonlinear patterns without requiring feature scaling and its ability to manage outliers effectively, and XGBoost for its proficiency in addressing class imbalance through weighted loss functions and its use of regularization to prevent overfitting.

Despite their strengths, both models present limitations, particularly in scenarios with extreme class imbalances (e.g., 2,823 severe to 46,409 non-severe pedestrian accidents). To mitigate this, we consider adjusting the class\_weight parameter and employing the Synthetic Minority Over-sampling Technique (SMOTE), which generates synthetic examples from the underrepresented class, ensuring a balanced dataset and enhancing model accuracy. The computational demands of these models increase with the complexity of their hyperparameter settings, which could impact overall performance. \\

The analysis is designed to target three specific outcomes: 

(1) whether an accident involves a pedestrian;

(2) whether an accident results in a serious injury or fatality;

(3) whether a serious accident involves a pedestrian.\\

For each target feature, we train both models and evaluate them using precision, accuracy, and ROC-AUC—metrics selected for their ability to capture both model correctness and discriminative power in identifying the most accurate outcomes. In addition, we examine feature importance results to identify risk factors associated with heightened pedestrian danger, providing actionable insights that can inform policy adjustments and improve road safety for pedestrians.

We divide the dataset into an 80\% training set and a 20\% test set, using the \texttt{stratify} parameter to ensure that class proportions in the training and test data mirror the original distribution. This strategy helps evaluate the models’ ability to generalize to new data. After splitting, we apply SMOTE to the training set to further address class imbalance, ensuring that the models are trained on a representative sample of the diverse scenarios present in the actual data.

\section*{3. Results Interpretation}
\vspace{-5pt} %
\subsection*{3.1 Spatial Analysis}

The spatial analysis component complements the predictive modeling by examining how pedestrian-related collision risks vary across geographic areas. After converting collision coordinates to a geospatial format and aligning them with the Local Authority District (LAD) boundary layer from the Office for National Statistics, each collision was assigned to its corresponding LAD through a spatial join. This procedure enabled the aggregation of collisions at the administrative level and provided a clearer representation of geographic concentration patterns in pedestrian-involved incidents.

The spatial join produced a near-complete match rate, indicating strong consistency between the STATS19 geocoded collision records and the LAD boundary dataset. District-level summaries revealed substantial geographic heterogeneity: some urban LADs exhibited markedly higher counts and shares of pedestrian-related collisions, while many rural districts showed lower absolute numbers but higher proportions relative to total collisions. This pattern aligns with established characteristics of the built environment. Densely populated and high-traffic urban areas tend to present more pedestrian exposure and more complex junction structures, leading to greater risk, whereas rural districts may experience fewer collisions overall but a higher likelihood of severe injury given speed conditions and road design.

To contextualize these spatial patterns, the analysis also considered the distribution of severe pedestrian collisions within LADs. Several districts exhibited elevated shares of severe outcomes relative to their total collision counts, suggesting that factors beyond traffic volume, such as road layout, lighting quality, enforcement intensity, or pedestrian infrastructure, may play an important role. These spatial differences underscore the need for geographically targeted interventions that address local risk profiles rather than relying solely on national policies.

\begin{figure}[ht]
    \centering
    \includegraphics[width=9.7cm]{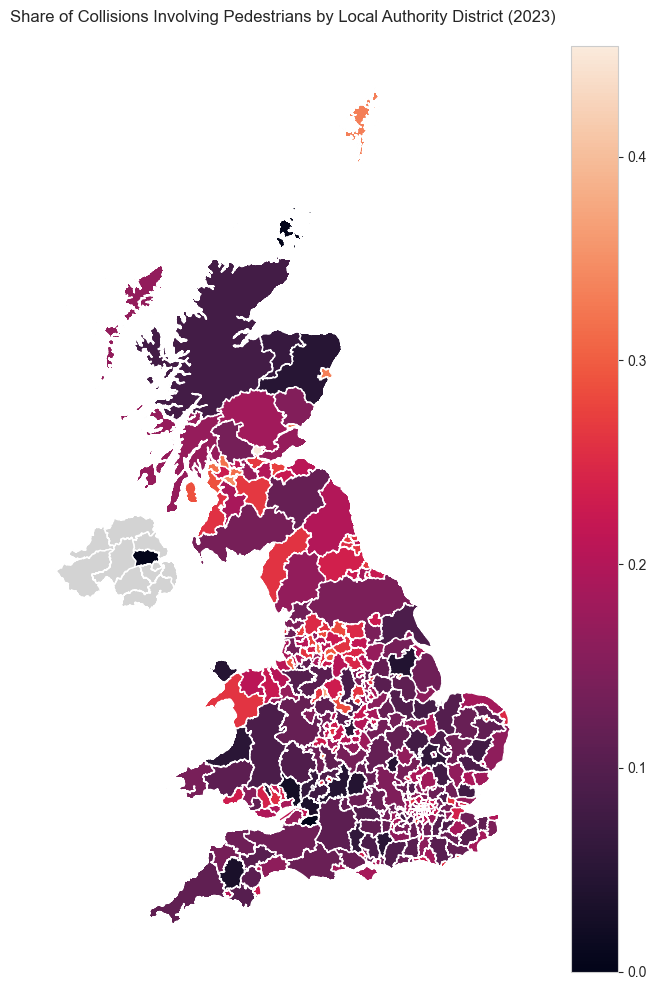}
    \caption{Choropleth Map of Pedestrian Collision Share by Local Authority District}
    \label{fig:map}
\end{figure}

\subsection*{3.2 Variable Importance Analysis}

\begin{table}[ht]
\centering
\caption{Accuracy of Models Across Different Targets}
\begin{tabular}{lccc}
\toprule
& & \multicolumn{2}{c}{Accuracy} \\
\cmidrule{3-4}
Model & Target Feature & Pre-tuning & After-tuning \\
\midrule
\multirow{3}{*}{Random Forest} & Pedestrian & 91.34\% & 91.71\% \\
& Severity Level & 65.24\% & 65.43\% \\
& Interaction & 87.80\% & 87.89\% \\
\midrule
\multirow{4}{*}{XGBoost} & Pedestrian & 91.89\% & 92.06\% \\
& Severity Level & 62.46\% & 62.70\% \\
& Interaction (original) & 84.93\% & 84.53\% \\
& Interaction (updated) & - & 87.81\% \\
\bottomrule
\end{tabular}
\end{table}

Our analysis targets three key aspects: pedestrian involvement, accident severity, and severe accidents involving pedestrians. For pedestrian involvement (Target 1), both Random Forest and XGBoost demonstrate high precision and recall for non-pedestrian-involved cases (class 0), underscoring their effectiveness in identifying more frequent scenarios. Notably, post-tuning improvements in XGBoost are substantial, enhancing its ability to handle imbalanced data and leading to improved precision and recall for pedestrian-involved accidents (class 1).\footnote{The complete classification reports for all six models, including all evaluation metrics, are provided in the Appendix.}

In predicting the severity of accidents (Target 2), both models achieve moderate success in non-severe cases but struggle with severe ones (class 1), even though tuning provides slight improvements. The inherent complexity of factors contributing to severe accidents results in lower accuracy and ROC-AUC values, highlighting the difficulty of this prediction task and the need for further refinement in feature selection, variable relevance, and model optimization.

The analysis of severe accidents involving pedestrians (Target 3) presents significant challenges, particularly in accurately predicting class 1 cases. Despite the high overall accuracy of around 90\% for both models, indicating effective class differentiation, performance remains limited due to the complexity of this category. The models struggle to achieve a balance between precision and recall, with hyperparameter optimization yielding only marginal improvements. Moreover, the need for distinct hyperparameter settings in XGBoost to attain higher accuracy further underscores the challenge of achieving robust performance, suggesting that high accuracy alone does not necessarily reflect a well-performing model in this context.

Our horizontal comparison indicates that post-tuning generally enhances model metrics such as F1-score, precision, and recall across both models for most classes and targets. XGBoost marginally outperforms Random Forest, particularly in handling more complex or imbalanced classes, owing to its advanced tuning capabilities and built-in mechanisms for class weight adjustment and regularization. The broader and more diverse hyperparameter space in XGBoost contributes to its superior performance, especially in class 1 predictions after tuning. However, under the objective of maximizing overall accuracy, the Random Forest model would be preferred for most targets.

As discussed in the model justification, predictive modeling of severe pedestrian accidents faces significant challenges due to the complexity of the problem and the pronounced imbalance in class distribution. These challenges are further compounded by the low precision and recall for class 1 predictions, even after applying techniques such as SMOTE to address class imbalance.

\subsection*{3.3 SHAP Analysis}

\begin{figure}[ht]
    \hspace*{-2cm} 
    \includegraphics[width=21cm]{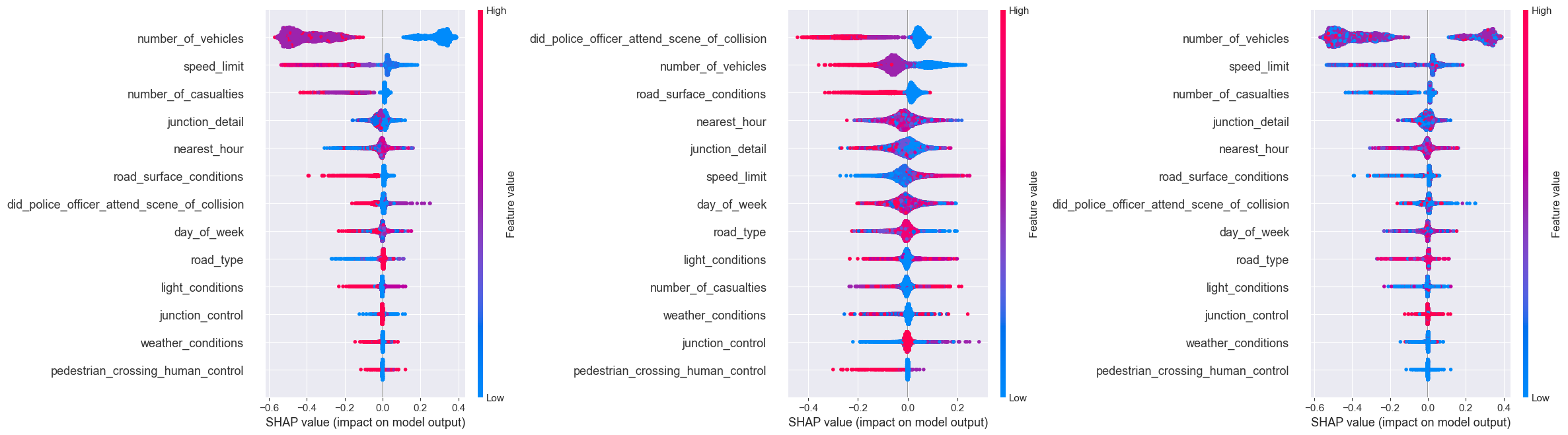}
    \caption{SHAP Plot of Random Forest Models}
    \label{fig:day_hour} 
\end{figure}

While accuracy differences between models were modest, the interpretability results from SHAP analysis reveal distinct patterns in feature influence, deepening understanding beyond performance metrics. The SHAP plots for the Random Forest and XGBoost models reveal the strong influence of features such as the number of vehicles and speed limit on accident severity predictions, with increases in either leading to higher severity classifications. Notably, features related to road surface and lighting conditions also affect predictions, as poor surface quality and nighttime settings are associated with more severe accidents. XGBoost exhibits a broader spread of influential features, reflecting its ability to capture complex interactions and its greater sensitivity to variations in feature values. For example, the prominence of police presence at the scene in XGBoost’s predictions suggests that more severe incidents are more likely to involve law enforcement intervention.

\begin{figure}[ht]
    \hspace*{-2cm} 
    \includegraphics[width=21cm]{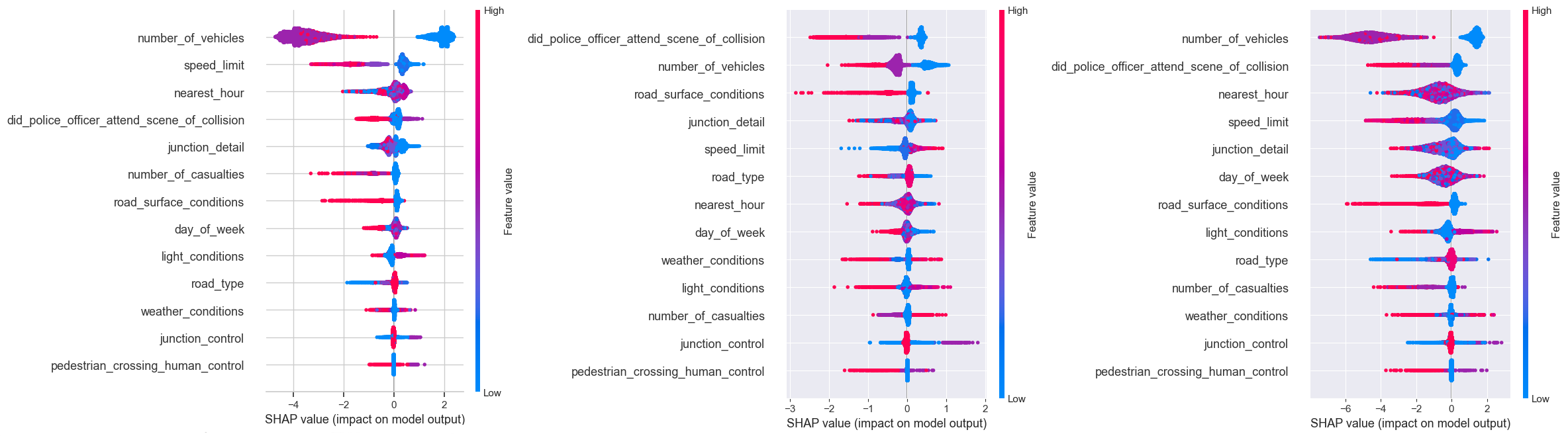}
    \caption{SHAP Plot of XGBoost Models}
    \label{fig:day_hour} 
\end{figure}

The insights from this study highlight the critical importance of model selection and hyperparameter tuning in managing complex, real-world datasets typical of road safety analysis. XGBoost, with its flexible tuning options and capacity to handle imbalanced classes, proves slightly more effective than Random Forest, particularly after hyperparameter adjustments that enhance model performance. These findings underscore the need to employ advanced machine learning techniques and iterative model optimization to improve predictive accuracy for rare but critical outcomes such as severe pedestrian accidents.

Furthermore, this study emphasizes the ongoing need for methodological innovation to strengthen predictive modeling in traffic safety applications, ensuring an appropriate balance between precision and recall through more comprehensive hyperparameter configurations. Maintaining this balance is essential in the context of road safety, where inaccuracies in predicting severe accidents can lead to substantial resource misallocation, either by over-predicting, which can waste valuable emergency response resources, or under-predicting, which may fail to prevent serious harm. Notably, when targeting collision severity levels, police presence emerges as one of the most influential factors. Therefore, optimizing the balance between model precision and recall through systematic hyperparameter tuning is vital for achieving more effective and efficient road safety management.

\section*{4. Conclusions}

This study of the 2023 UK Road Safety Data underscores the critical role of advanced analytical methods in understanding the conditions associated with pedestrian-involved collisions and injury severity. By integrating machine learning models with spatial analysis, the study provides a multifaceted perspective on pedestrian safety that goes beyond descriptive statistics or single-method approaches. The predictive modeling results demonstrate that XGBoost, with its flexible tuning capacity and stronger handling of imbalanced outcomes, offers modest yet consistent performance advantages over Random Forest, especially following hyperparameter optimization. SHAP analysis complements these findings by identifying key factors that influence accident severity, such as the number of vehicles involved, prevailing speed limits, lighting conditions, and road surface quality. These results highlight the complexity of pedestrian safety and the importance of both environmental and behavioral factors in shaping collision outcomes.

The spatial analysis further enriches the interpretation of these results by revealing the pronounced geographic variation in pedestrian risks across Local Authority Districts. Urban areas exhibit higher concentrations of pedestrian collisions, while certain rural areas demonstrate disproportionately severe outcomes relative to their total number of crashes. These patterns suggest that local infrastructure, traffic density, enforcement practices, and environmental conditions significantly influence both the likelihood and severity of pedestrian accidents. The combined methodological approach therefore provides a more comprehensive understanding of pedestrian safety and supports more targeted, context-specific policy interventions rather than broad national strategies alone.

Taken together, the findings emphasize the value of integrating data-driven predictive modeling with geographic context to improve the precision of policy recommendations. Interventions such as speed management, enhanced lighting in high-risk zones, investment in pedestrian infrastructure, and targeted enforcement may be informed by both the location-specific collision patterns and the feature importance identified in the models. Furthermore, the analysis illustrates the importance of transparent and interpretable machine learning tools, such as SHAP interpretation, in bridging technical insights with practical decision-making for transportation safety authorities.

\subsection*{4.1 Implications for Policy Adjustments}

The findings of this study provide actionable insights for strengthening pedestrian safety through data-informed policy development and targeted infrastructural improvements. The strong influence of speed limit, lighting conditions, and road surface quality on collision severity indicates that stricter speed management policies, particularly in high-risk zones or areas with vulnerable pedestrian populations, may help reduce the likelihood of severe outcomes. Enhancing roadway infrastructure through improved lighting, clearer signage, surface repairs, and redesigned junctions can further mitigate collision risks by addressing environmental conditions identified as influential in the predictive models.

The spatial analysis emphasizes the geographic variability of pedestrian risk, suggesting that local authorities should incorporate district-level collision patterns into their planning and resource allocation. Predictive analytics can support this process by identifying locations and times where risks are elevated, allowing for more efficient deployment of enforcement personnel, targeted public awareness initiatives, and optimized emergency response strategies.

Integrating these analytical insights into transport governance frameworks enables policymakers to transition from reactive safety measures to proactive, evidence-based interventions. Such an approach addresses both environmental and behavioral dimensions of road safety, providing a strong foundation for coordinated, multi-sector strategies. These strategies, along with opportunities for methodological enhancement, are explored in the subsequent section on future research.

\subsection*{4.2 Further Research and Directions}
While this study provides foundational insights into pedestrian safety, further research is needed to expand the range of variables incorporated into predictive models and to deepen understanding of the mechanisms driving road traffic collisions. Temporal patterns in pedestrian movement, variations in pedestrian-vehicle interactions, and broader environmental conditions may significantly influence collision likelihood yet fall outside the scope of the current dataset. For example, the study by Gerogiannis and Bode on pedestrian behavior at unmarked crossings shows that local behavioral norms and crossing practices play a central role in shaping accident risk. These findings suggest that incorporating behavioral heterogeneity and localized movement patterns into future models could substantially improve both explanatory power and policy relevance.

Environmental features also warrant additional investigation. The analysis by Marchant, Hale, and Sadler, which identifies a positive association between increased lighting and accident frequency, highlights the need to explore how environmental interventions interact with driver and pedestrian behavior as well as with underlying infrastructural constraints. This complexity points to the value of interdisciplinary approaches that combine data-driven modeling with perspectives from urban planning, human factors research, and public health. Integrated methods that draw upon these fields may yield a more comprehensive understanding of how environmental characteristics influence road safety outcomes.

Methodologically, it is important to distinguish between the predictive modeling used in this study and causal inference. While our SHAP analysis identifies features that are strongly associated with accident severity, these results represent correlations rather than causal effects. Unlike econometric methods such as Difference-in-Differences (DiD), this cross-sectional analysis of 2023 data cannot isolate the specific impact of the 2022 Highway Code amendments. Future research could employ a panel data approach to rigorously evaluate the causal effectiveness of these policy changes.

Finally, Staton et al.'s (2022) critique of the recent stagnation in reducing road fatalities across the United Kingdom underscores the importance of institutional capacity, governance structures, and implementation strategies. The effectiveness of data-informed road safety interventions depends not only on model accuracy but also on the ability of local and national authorities to enact, enforce, and sustain evidence-based policy measures. Future research should, therefore, incorporate governance and community-level factors, examining how institutional readiness, resource allocation, and public engagement influence the success of road safety policies. Advancing pedestrian safety requires a holistic analytical framework that integrates empirical modeling with an understanding of policy processes, community dynamics, and sustained commitment to long-term safety improvements.

\subsection*{4.3 Limitations and Challenges}

Hyperparameter tuning is a critical component of machine learning model development, yet it presents multiple methodological and computational challenges. As demonstrated in the implementation of the Random Forest and XGBoost models, identifying optimal hyperparameter combinations is both time consuming and computationally intensive. The tuning process conducted in this study, although systematic, may not have been exhaustive enough to achieve peak performance for all models and target variables. In several instances, tuned models performed similarly or even slightly worse than their baseline counterparts, which may reflect limitations in the search space, variability in model sensitivity, or constraints imposed by computational resources. Additional grid search experiments with expanded parameter ranges produced minor improvements but yielded inconsistent results across different prediction targets, indicating that more sophisticated tuning procedures may be required. Furthermore, the need for target-specific threshold adjustments complicates direct model comparisons and highlights the complexity of modeling highly imbalanced safety outcomes.

A second limitation stems from the characteristics of the underlying dataset. The collision and casualty records from the UK Department for Transport include only police-reported personal injury collisions. This reporting mechanism introduces the possibility of under-reporting and systematic omissions. Collisions without injuries, as well as incidents resolved privately without police involvement, are excluded. Some cases may also go unreported due to a lack of awareness of reporting requirements or intentional avoidance by drivers involved in violations. Consequently, the dataset may underestimate the true scale and distribution of road collisions, which could affect the generalizability of the study’s findings.

In addition, the structure of the casualty data presents challenges when attempting to incorporate sociodemographic variables. Individual casualties are recorded separately even when they originate from the same collision, resulting in repeated collision identifiers and complicating efforts to align person-level characteristics with collision-level features. The absence of reliable and complete sociodemographic information limits the ability to examine equity dimensions or vulnerable population groups within the collision dataset. This constraint reduces the scope for analyzing important social determinants of road safety risk.

Addressing these limitations will require integrating complementary data sources such as hospital admissions, emergency medical service records, or insurance claims to improve coverage and reduce reporting bias. Employing advanced imputation techniques and conducting sensitivity analyses may help mitigate the impact of missing or incomplete information. More efficient hyperparameter optimization strategies, including Bayesian optimization and automated search frameworks, may improve both computational efficiency and predictive performance. Combining these approaches would strengthen future analyses and enhance the robustness, interpretability, and policy relevance of predictive models in road safety research.

\newpage
\section{Appendix}

\begin{table}[!htbp] 
\centering
\caption{Evaluation of Pedestrian Predictive Models}
\begin{tabular}{lcccccc}
\toprule
& \multicolumn{2}{c}{Random Forest Model} & \multicolumn{2}{c}{XGBoost Model} \\
\cmidrule(lr){2-3} \cmidrule(lr){4-5}
Metric & Pre-tuning & Post-tuning & Pre-tuning & Post-tuning \\
\midrule
\textbf{Class(0)} & & & & \\
Precision & 0.97 & 0.97 & 0.97 & 0.98 \\
Recall & 0.92 & 0.93 & 0.93 & 0.93 \\
F1-Score & 0.95 & 0.95 & 0.95 & 0.95 \\
Support & 8071 & 8071 & 8071 & 8071 \\
\midrule
\textbf{Class(1)} & & & & \\
Precision & 0.72 & 0.72 & 0.72 & 0.73 \\
Recall & 0.86 & 0.87 & 0.89 & 0.90 \\
F1-Score & 0.78 & 0.79 & 0.80 & 0.80 \\
Support & 1776 & 1776 & 1776 & 1776 \\
\midrule
Accuracy & 91.34\% & 91.71\% & 91.89\% & 92.10\% \\
ROC\_AUC & 0.940 & 0.944 & 0.953 & 0.955 \\
\bottomrule
\end{tabular}
\end{table}


\begin{table}[!htbp] 
\centering
\caption{Evaluation of Severity Level in Predictive Models}
\begin{tabular}{lcccccc}
\toprule
& \multicolumn{2}{c}{Random Forest Model} & \multicolumn{2}{c}{XGBoost Model} \\
\cmidrule(lr){2-3} \cmidrule(lr){4-5}
Metric & Pre-tuning & Post-tuning & Pre-tuning & Post-tuning \\
\midrule
\textbf{Class(0)} & & & & \\
Precision & 0.80 & 0.80 & 0.83 & 0.83 \\
Recall & 0.73 & 0.73 & 0.64 & 0.64 \\
F1-Score & 0.76 & 0.76 & 0.72 & 0.73 \\
Support & 7538 & 7538 & 7538 & 7538 \\
\midrule
\textbf{Class(1)} & & & & \\
Precision & 0.31 & 0.32 & 0.32 & 0.33 \\
Recall & 0.41 & 0.41 & 0.56 & 0.57 \\
F1-Score & 0.36 & 0.36 & 0.41 & 0.42 \\
Support & 2309 & 2309 & 2309 & 2309 \\
\midrule
Accuracy & 65.24\% & 65.43\% & 62.46\% & 62.70\% \\
ROC\_AUC & 0.621 & 0.621 & 0.625 & 0.659 \\
\bottomrule
\end{tabular}
\end{table}

\begin{table}[ht]
\centering
\caption{Evaluation of Interaction Models}
\begin{tabular}{lcccccc}
\toprule
& \multicolumn{2}{c}{Random Forest Model} & \multicolumn{2}{c}{XGBoost Model} \\
\cmidrule(lr){2-3} \cmidrule(lr){4-5}
Metric & Pre-tuning & Post-tuning & Pre-tuning & Post-tuning \\
\midrule
\textbf{Class(0)} & & & & \\
Precision & 0.97 & 0.97 & 0.98 & 0.96 \\
Recall & 0.90 & 0.90 & 0.86 & 0.90 \\
F1-Score & 0.93 & 0.93 & 0.91 & 0.93 \\
Support & 9282 & 9282 & 9282 & 9282 \\
\midrule
\textbf{Class(1)} & & & & \\
Precision & 0.23 & 0.23 & 0.23 & 0.22 \\
Recall & 0.47 & 0.49 & 0.67 & 0.46 \\
F1-Score & 0.31 & 0.32 & 0.34 & 0.30 \\
Support & 565 & 565 & 565 & 565 \\
\midrule
Accuracy & 87.80\% & 87.89\% & 84.93\% & 87.81\% \\
ROC\_AUC & 0.849 & 0.859 & 0.866 & 0.838 \\
\bottomrule
\end{tabular}
\end{table}

\begin{table}[ht]
\centering
\caption{Comparison between Different Hyperparameter Settings}
\begin{tabular}{lcc}
\toprule
& \textbf{XGBoost (Original setting)} & \textbf{XGBoost (New setting)} \\
\midrule
\textbf{Class(0)} & & \\
Precision & 0.98 & 0.96 \\
Recall & 0.86 & 0.90 \\
F1-Score & 0.91 & 0.93 \\
Support & 9282 & 9282 \\
\midrule
\textbf{Class(1)} & & \\
Precision & 0.22 & 0.22 \\
Recall & 0.68 & 0.46 \\
F1-Score & 0.33 & 0.30 \\
Support & 565 & 565 \\
\midrule
Accuracy & 84.53\% & 87.81\% \\
ROC\_AUC & 0.8672\% & 0.8382\% \\
\bottomrule
\end{tabular}
\label{tab:hyperparameter_comparison}
\end{table}

\begin{figure}[ht]
    \centering 
    \includegraphics[width=17cm]{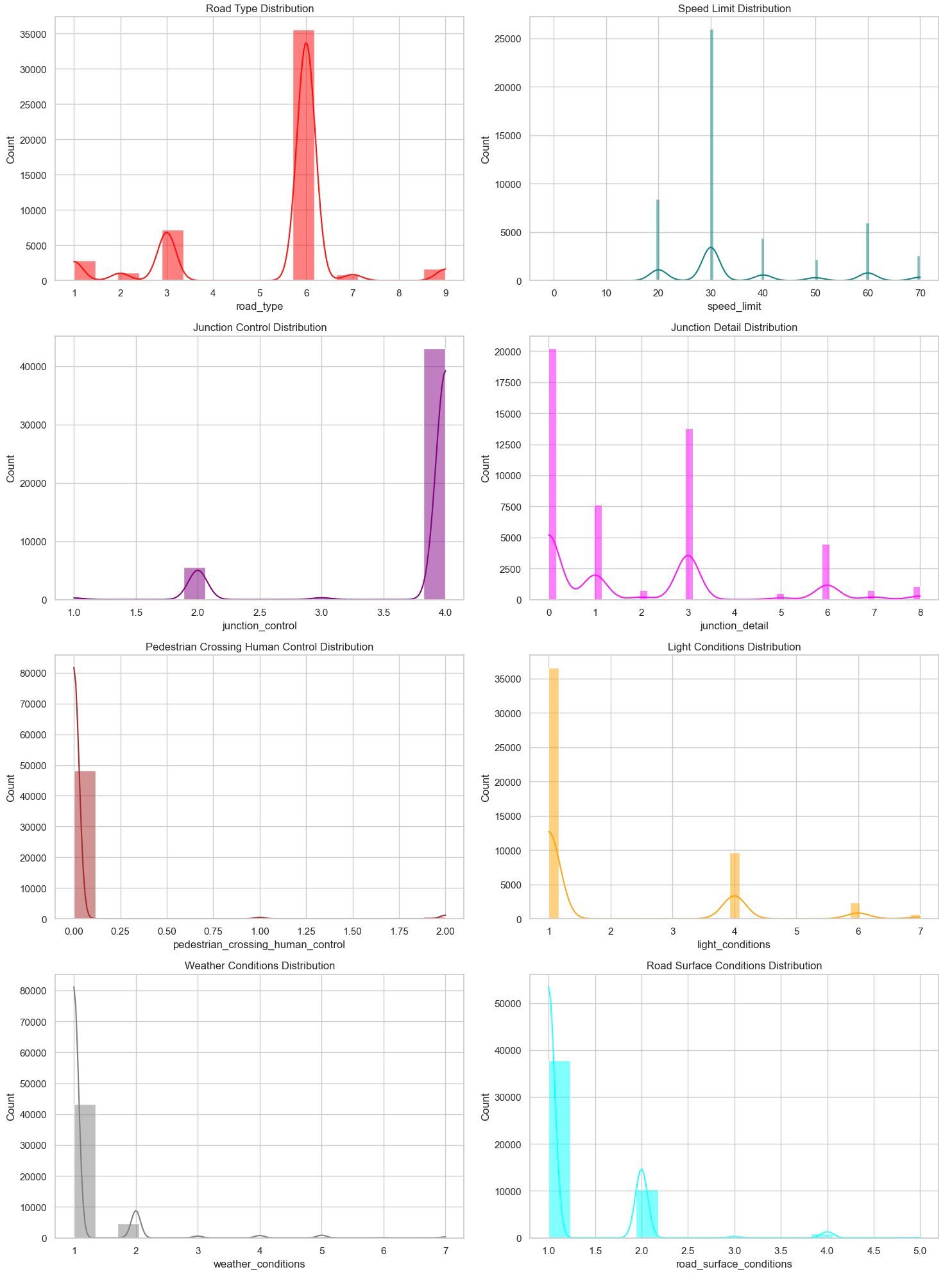}
    \caption{Distribution Plots for Remaining Variables}
    \label{fig:remain} 
\end{figure}

\vspace{10pt} %

\end{document}